\documentclass[letterpaper, 10 pt, conference]{ieeeconf} 
\usepackage[utf8]{inputenc}
\usepackage{cite}
\usepackage{amsmath,amssymb,amsfonts}
\usepackage{algorithmic}
\usepackage{graphicx}
\usepackage{textcomp}
\usepackage{gensymb}
\usepackage{array,multirow}
\usepackage{soul}
\setstcolor{blue}
\usepackage{diagbox}
\usepackage{booktabs}
\usepackage{threeparttable}

\def\BibTeX{{\rm B\kern-.05em{\sc i\kern-.025em b}\kern-.08em
    T\kern-.1667em\lower.7ex\hbox{E}\kern-.125emX}}
\usepackage{color, colortbl}
\definecolor{Gray}{gray}{0.9}
\definecolor{White}{gray}{1}

\definecolor{bgreen}{rgb}{0.0, 0.26, 0.15}
\definecolor{brickred}{rgb}{0.8, 0.25, 0.33}
\definecolor{iceberg}{rgb}{0.44, 0.65, 0.82}
\definecolor{indigo(dye)}{rgb}{0.0, 0.25, 0.42}
\definecolor{junebud}{rgb}{0.74, 0.85, 0.34}
\definecolor{lapislazuli}{rgb}{0.15, 0.38, 0.61}
\definecolor{lightpastelpurple}{rgb}{0.69, 0.61, 0.85}
\definecolor{mediumslateblue}{rgb}{0.48, 0.41, 0.93}

\newcommand*{\rom}[1]{\uppercase\expandafter{\romannumeral #1\relax}}
\IEEEoverridecommandlockouts        

\title{\LARGE \bf 
A Focus of Attention–Based Virtual Training Platform for Pre-Prosthetic Myoelectric Skill Acquisition: A Proof-of-Concept Study}

\author{$^*$Xiaochen Zhang$^{1,2}$ and Sigrid Dupan$^{1,2,3}$%
\thanks{*This work was generously funded by the Taighde Éireann—Research Ireland under Grant Number 21/PATH-S/9605.}
\thanks{*Corresponding author is Xiaochen Zhang, xiaochen.zhang@ucd.ie.}
\thanks{$^{1}$ School of Electrical and Electronic Engineering, University College Dublin,
Dublin, Ireland.}
\thanks{$^{2}$ Centre for Biomedical Engineering, University College Dublin, Dublin, Ireland.}
\thanks{$^{3}$ School of Mechanical and Materials Engineering, University College Dublin, Dublin, Ireland.}
\thanks{©2026 IEEE. Personal use of this material is permitted. Permission from IEEE must be obtained for all other uses, in any current or future media, including reprinting/republishing this material for advertising or promotional purposes, creating new collective works, for resale or redistribution to servers or lists, or reuse of any copyrighted component of this work in other works.}
}
\begin{document}

\maketitle

\thispagestyle{empty}
\pagestyle{empty}

\begin{abstract}
Advances in myoelectric prosthetic technology have substantially increased the functional potential of modern devices. Accordingly, heightened control demands have led to the acknowledgement of pre-prosthetic training as a key stage in the acquisition of myoelectric skills.
Existing training paradigms largely emphasize internal muscle activation, while external, goal-directed outcomes required for effective real-world use are often neglected.
We address this gap by introducing a virtual pre-prosthetic training platform that integrates EMG-driven cursor with animated hand gestures, enabling the delivery of both muscle‑level and functional‑level feedback. In this proof‑of‑concept study, participants were assigned to one of two focus of attention (FoA) protocols, each incorporating both feedback types but differing in whether internal or external FoA was emphasised.
Participants successfully acquired and retained myoelectric skill across both protocols, but distinct performance characteristics and learning strategies emerged, indicating that both FoAs contribute meaningfully to learning and that their timing may play an important role. External FoA was positively associated with retention, suggesting that it may strengthen the link between training and skill acquisition.
Together, the results demonstrate the feasibility of an FoA‑based virtual training platform for pre‑prosthetic applications and indicate that it can provide a foundation for designing training protocols that better prepare users for prosthetic use.

\end{abstract}


\IEEEpeerreviewmaketitle

\section{introduction}

Upper-limb loss affects millions of people worldwide, and many rely on prosthetic devices to restore functional independence in daily activities \cite{shahsavari2020upper,bates2020technological}. Over the past decade, substantial advances have been achieved in prosthetic design and device functionality, including increased degrees of freedom, improved actuation, and enhanced sensing capabilities \cite{boccardo2023development,trent2020narrative,gu2023soft,furui2019myoelectric,capsi2025merging}.
In parallel, myoelectric control has become widely adopted. Myoelectric prostheses operate by recording electromyography (EMG) signals from residual forearm muscles to estimate the user’s motor intent and translate it into corresponding prosthetic movements \cite{godfrey2018softhand}. 
While these advances have significantly expanded the functional capabilities of upper-limb prostheses, effective use remains highly challenging and relies heavily on the reliable and precise modulation of residual muscle activity.

Targeted prosthetic training shows evidence in improving functional outcomes and promoting long-term device acceptance \cite{bouwsema2014changes}. In particular, pre-prosthetic training enables early intervention during the period between amputation and the fitting of a wearable prosthesis and also helps reduce the risk of muscle atrophy \cite{chappell2022virtual}. Virtual prosthetic training tools are well-suited for this stage, as they allow users to practice control strategies without imposing mechanical load or excessive strain on the residual limb or healing wound \cite{garske2023enhancing}. Such early training is especially important for myoelectric prostheses, as users must learn to associate EMG signals with functional outcomes and develop stable sensorimotor mappings before interacting with a real prosthetic device \cite{resnik2018evaluation,farag2025myoelectric}.

Conventional myoelectric prosthetic training paradigms, whether based on direct control or pattern recognition, often emphasize explicit control of muscle contractions, directing the user’s attention internally toward muscle‑level movement, typically through EMG-based feedback \cite{gottwald2020internal}. 
These approaches are effective for developing consistent muscle modulation and early muscle‑specific control \cite{kristoffersen2019effect}, and they can support the skill transfer under certain conditions, although such training paradigms often lead to slower initial learning \cite{dupan2025successful}. This highlights a broader challenge in prosthesis training: identifying an optimal training strategy that balances eventual learning with speed and an engaging learning experience. 

One promising direction is to shift from an internal to an external focus of attention (FoA), where training becomes increasingly centered on functional outcomes and their effects on the environment \cite{parr2022scoping}. This aligns with motor learning theory, which distinguishes between an internal FoA, concerned with how movements are produced, and an external FoA, directed toward the consequences of movement \cite{wulf2013attentional}. 
Myoelectric games (myogames) offer a promising means to incorporate motor learning theory into prosthesis training by shifting the user's focus from muscle activation itself to observable outcomes of their actions \cite{kristoffersen2020serious, kristoffersen2021user}. 
By embedding EMG control within goal-oriented and interactive tasks, myogames can promote an external FoA, enhance user engagement, and support a deeper understanding of the influence of muscle activation on control outcomes \cite{prahm2017increasing, kristoffersen2020serious}.

However, evidence from motor learning research highlights that task similarity, the extent to which training tasks resemble real-life activities, is a key factor in facilitating effective skill acquisition \cite{sebastian2016quantifying, proteau1998practice}; 
The output of current myogames often shows reduced similarity between the game effect and functional prosthetic tasks \cite{tabor2017quantifying}, potentially limiting their relevance for functional prosthesis use \cite{van2016learning, maas2024assessing}. 
This suggests a need for training platforms that intentionally integrate the feedback with external FoA while maintaining functional relevance.

This paper presents a proof-of-concept study of a FoA–based virtual training platform for pre-prosthetic myoelectric skill acquisition. 
The proposed platform integrates feedback from the EMG-driven cursor and animated hand gesture, which simultaneously support internal and external FoA while maintaining task similarity to future prosthetic use. 
The key contributions of this work are twofold:
1) the development of a virtual myoelectric training platform that intentionally combines internal and external FoA cues within one training environment; and
2) a preliminary evaluation of the feasibility of this FoA–based training approach for supporting myoelectric skill acquisition in a pre-prosthetic context under two FoA protocols.

\section{materials and methods}

\subsection{Training platform design}

\subsubsection{EMG}
Two wireless EMG sensors (Trigno, Delsys, Inc., USA) were placed on the flexor carpi radialis (FCR) and extensor carpi radialis (ECR) muscle groups to capture the muscle activities. 
The measured EMG signals were sampled at 2000 Hz and band-pass filtered between 10 Hz and 500 Hz. 

To enable subject-specific normalization of EMG signals, a calibration procedure was conducted prior to the first day of training. 
During this procedure, mean absolute values (MAVs) of the EMG signals were recorded during muscle relaxation and during a comfortable contraction. MAVs were computed using a 750 ms sliding window updated every 20 ms. 
Specifically, participants were instructed to first fully relax their muscles to obtain resting MAV values, and then to perform a comfortable, repeatable contraction that could be sustained without fatigue to obtain contraction MAV values. 

These two reference values served as the basis for subsequent EMG processing, with each channel's signal being normalized in real time according to

\begin{equation}
\hat{y} = \frac{y - y_r}{y_c - y_r}
\end{equation}

with $\hat{y}$ the normalized muscle activity, $y$ the MAV calculated over the current 750 ms window, $y_r$ the MAV during muscle relaxation, and $ y_c$ the MAV during a comfortable muscle contraction.

\subsubsection{Myoelectric control interface (MCI)}

\begin{figure}[htbp]
	\centering
	\includegraphics[width=0.48\textwidth]{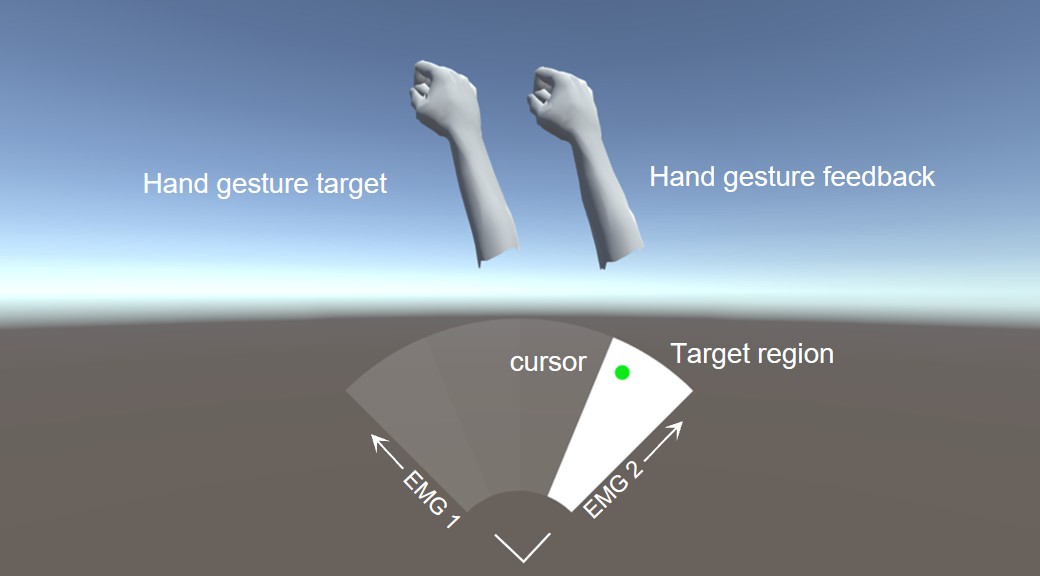}
	\caption{Myoelectric control interface (MCI). The interface integrates a V-shaped EMG cursor workspace (lower) and an EMG-driven virtual hand model (upper). Normalized EMG signals from two muscle channels (EMG 1 and EMG 2) control the position of the green cursor within the V-shaped workspace, with the highlighted sector indicating the target region. In parallel, the same EMG inputs drive the virtual hand gesture feedback (right) toward a predefined hand gesture target (left).
	}
	\label{mci}
\end{figure}

The MCI was developed as a custom application using C\# and the Unity engine (Unity Technologies, USA), as illustrated in Fig. \ref{mci}. 
The MCI provides two visual feedback strategies designed to support different focuses of attention: (i) a two-dimensional V-shaped EMG cursor interface and (ii) an EMG-driven virtual hand model. 
Both were controlled in real time by the normalized EMG signals, as described in the previous section, which were streamed from a Python-based processing pipeline using the AxoPy library \cite{lyons2019axopy} to the Unity environment via a TCP/IP protocol.

In the lower portion of the MCI, the V-shaped interface provides explicit visual feedback of muscle activation, representing an internal FoA. 
It follows the same motor learning-based principles established in \cite{dyson2018myoelectric, stuttaford2023delaying,dupan2025successful}, and allows for abstract control where participants learn to modulate contractions of the FCR and ECR muscles to control the position of the green cursor within a two-dimensional V-shaped workspace comprising four predefined target regions. These targets can then be mapped onto any grasp of a prosthesis \cite{dupan2025successful}.
The cursor’s position along the V-shaped axes is determined by the normalized EMG amplitude ($\hat{y}$) of each muscle, with the axes extremities defined at $\hat{y} = 1$.
Each of the four predefined target regions is bounded by a minimum activation threshold of $\hat{y} = 0.3$, requiring participants to generate sufficient muscle activity to reach and remain within the target.
In each trial, one of the four regions from the V-shaped workspace was highlighted, and participants modulated their muscles to navigate the cursor into the target region. Each trial provided 0.75 s for reaching the target, followed by a 1.5 s hold period.

\begin{figure*}[htbp]
	\centering
	\includegraphics[width=0.75\textwidth]{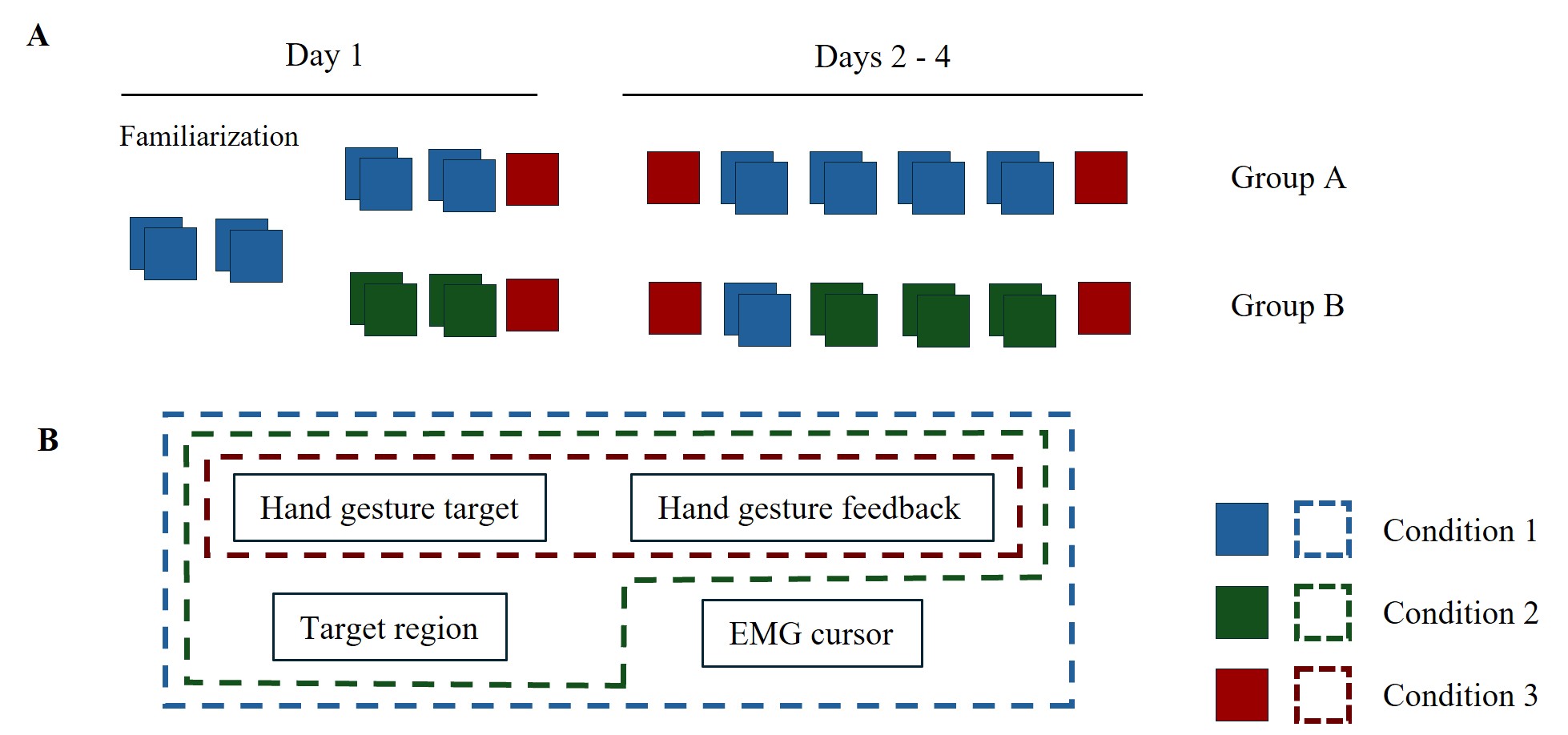}
	\caption{Overview of the experimental protocol. 
    A. Trial block structure experienced by participants across the four experimental days. 
    Day 1 consisted of a familiarization session common to all participants. During the second half of Day 1 and Days 2–4, participants were divided into two groups, each following a different training plan with distinct FoA conditions across training blocks. 
    B. Structure of the three conditions. 
    Each dashed box represents one condition and illustrates the combination of target and visual feedback provided.
	}
	\label{protocol}
\end{figure*}

In parallel, two virtual hand models were displayed in the upper portion of the screen. The models used a boned-and-rigged architecture adapted from the framework established by Boschmann et al. \cite{boschmann2021immersive}, enabling the rendering of complex, realistic grasping animations. 
Four hand gesture animations, “index", “pincer", “lateral", and “power", were mapped, from left to right, onto the four target regions of the V‑shaped workspace, while a relaxed muscle state corresponded to an “open" gesture.
During each trial, the left virtual hand model presented the target gesture associated with the highlighted region. The right model was updated continuously based on the user’s EMG signals and displayed the gesture corresponding to the region the cursor entered. Together, the two hand models provide immediate visual feedback on whether the muscles were correctly activated and matched the intended target gesture.

Furthermore, this MCI was designed as a flexible training platform in which either interface component, the lower V‑shaped control workspace or the upper virtual hand display, can be used independently or in combination, allowing the system to accommodate different training formats with a distinct FoA.

\subsection{Participants}

Six limb-intact participants (sex: 3 female/3 male, age: 30.0 ± 4.2) free from neurological or motor disorders volunteered for this study. 
Ethical approval was granted by the Human Research Ethics Committee at University College Dublin (UCD) – Sciences (LS-24-78-Dupan). 
The research was conducted in accordance with the principles of the Declaration of Helsinki, and all participants provided written informed consent.

\subsection{Experimental protocol}

The experiment was conducted in the Neuromuscular Systems Lab at UCD. During training, participants sat at a viewing distance of 1 meter from a 24-inch screen displaying the MCI task, and were instructed to flex their elbows to approximately 90° while keeping their wrists in a neutral position.
Each participant completed a 4-day training protocol, with one training session per day lasting approximately one hour. The participants were randomly assigned to two groups, Group A or Group B, with distinct experimental protocols for both groups, as shown in Fig. \ref{protocol}A.
The experiment employed myoelectric control tasks under three different focus-of-attention conditions, as shown in Fig. \ref{protocol}B:

\begin{itemize}
\item Condition 1: Participants were provided with a target region within the V-shaped workspace, real-time EMG cursor position, the virtual hand gesture target, and virtual hand gesture feedback. This condition allowed participants to simultaneously attend to both muscle activation patterns and their associated functional movement outcomes.

\item Condition 2: Participants were provided with the target within the V-shaped workspace, the virtual hand gesture target, and virtual hand gesture feedback, but without visual feedback of the EMG cursor position. By withholding direct visualization of muscle activity, this condition encouraged an external FoA.

\item Condition 3: Participants interacted solely with the virtual hands (target and feedback), without any explicit visual feedback related to muscle activation or EMG cursor position.

\end{itemize}

On Day 1, participants were first introduced to the fundamental principles of myoelectric control and the virtual training environment. Participants then completed 160 trials (4 blocks of 40 trials each) to familiarize themselves with the MCI, to understand available visual feedback and task targets before entering the training phase.

Following the familiarization session, participants were randomly assigned to one of two groups, Group A or Group B, each following a distinct training protocol throughout the rest of Day 1 and Days 2–4.
Training was organized into blocks of 40 trials, with participants completing 4 training blocks on Day 1 and 8 blocks on each of Days 2-4. 
Group A was trained under Condition 1, in which both feedback of the target muscle contraction and the corresponding output hand gesture were provided. 
Group B followed an alternative protocol that directed participants to shift their attention to external, task-relevant cues from the virtual hand model: on Days 2–4, 6 training blocks were conducted under Condition 2 in place of Condition 1.

Condition 3 was defined as a retention test and used at the beginning and end of each training day for both groups, with the exception of the start of day 1. These blocks were used to evaluate skill-acquisition performance.

After each training day, participants were asked to quantify where their attention was during the tasks, specifically what percentage was on the virtual hand models.

\subsection{Measures}

A score was calculated for each trial, defined as the percentage of time that the cursor remained within the target region, or the percentage of time during which the right virtual hand model correctly matched the gesture performed by the left hand, throughout the hold period. 
Following this principle, two performance-based metrics were derived: (1) training scores, computed across the training blocks, and (2) retention scores, computed across the test blocks.

In addition to these performance measures, the self-reported attentional allocation from both groups after each training day, specifically, what percentage was focused on the virtual hand models, was summarized.

To investigate factors associated with retention performance, correlation analyses were conducted between retention scores and two training-related measures: training performance and attention to the virtual hand. 
For each training day, training performance was defined as the performance in the last training block preceding the test block, attention to the virtual hand was defined as the self-reported attention score, and retention performance was defined as performance in the final test block of the day. 
Correlations were computed separately for Groups A and B, and statistical significance was assessed for each relationship.

\section{Results}

\begin{figure}[htbp]
	\centering
	\includegraphics[width=0.45\textwidth]{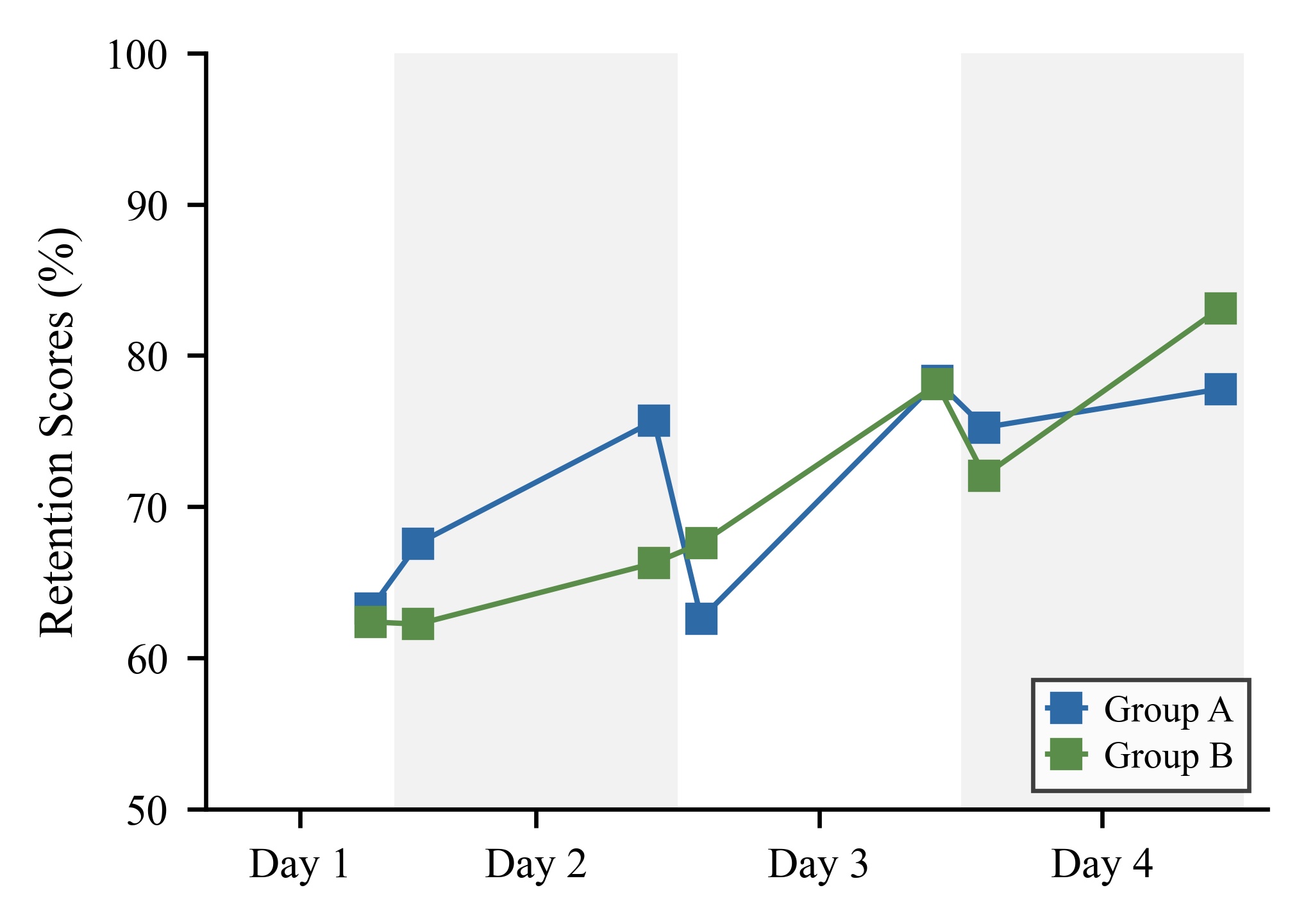}
	\caption{Retention performance for Groups A and B across the four training days. 
    Mean retention scores (\%) were calculated for each group at each retention block.
	}
	\label{retention}
\end{figure}

Retention scores across the four training days for participants in Groups A and B are shown in Fig. \ref{retention}. 
Group-level mean retention scores were calculated for each group at each time point.
Both groups began at comparable levels, as measured at the end of Day 1 (Group A: 63.27\%; Group B: 62.38\%), and showed an overall improvement in retention performance over time, although scores tended to decline slightly between consecutive training days.
By the final training day, both groups achieved comparable retention levels, with slightly higher scores observed in Group B (Group A: 77.79\%; Group B: 83.18\%).

\begin{figure}[htbp]
	\centering
	\includegraphics[width=0.45\textwidth]{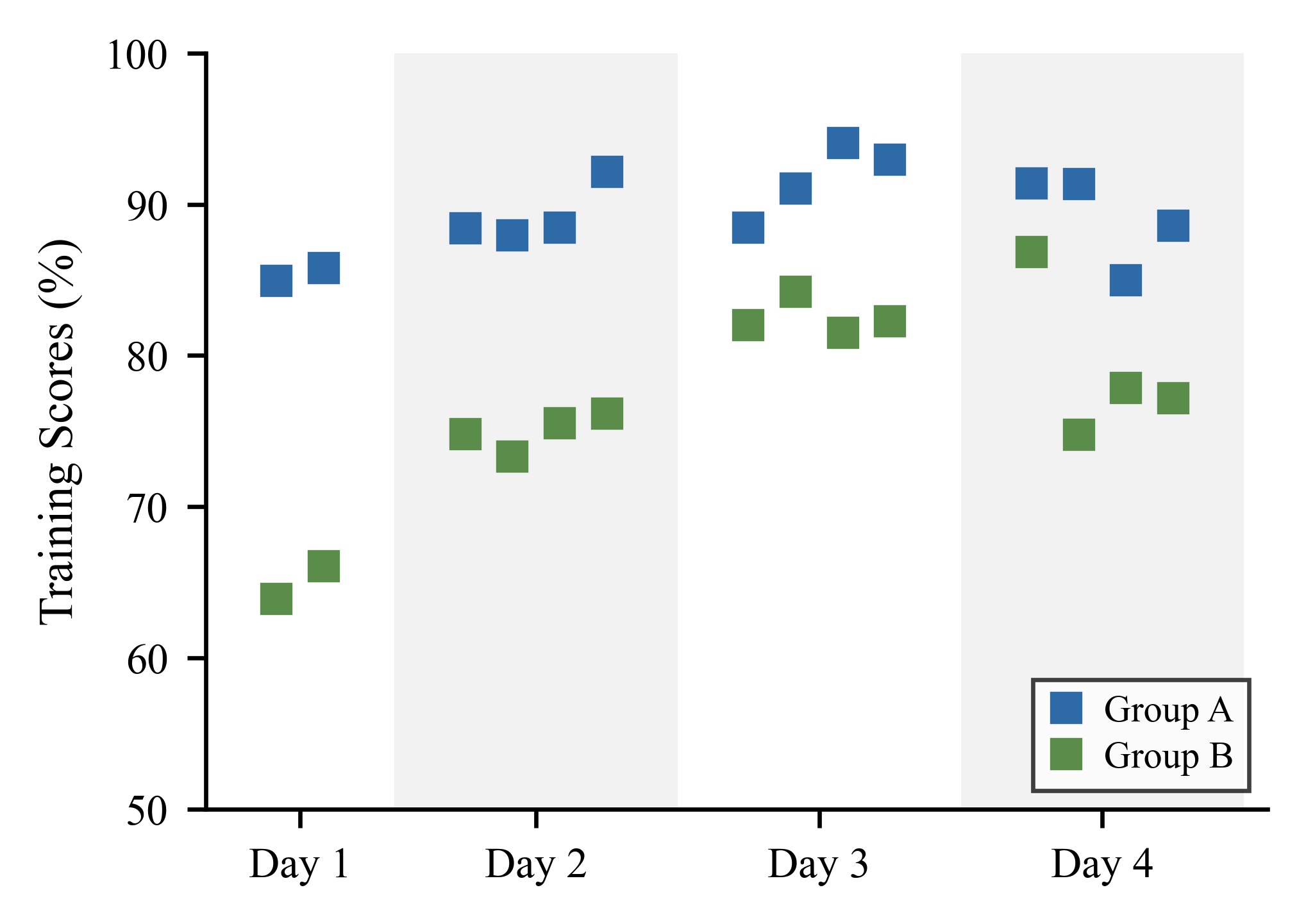}
	\caption{Training scores for Groups A and B across the four training days. 
    Mean scores were calculated from the 2nd, 4th, 6th, and 8th training blocks. 
    Group A trained under Conditions 1, while Group B trained under Conditions 1 and 2, as shown in Fig. \ref{protocol}.
	}
	\label{training}
\end{figure}

Fig. \ref{training} shows training scores across the four training days for participants in Groups A and B. Group‑level mean scores were computed at each time point using the 2nd, 4th, 6th, and 8th training blocks. 
Across all four days, Group A consistently achieved higher training scores than Group B. This pattern was evident both in the initial blocks of Days 2–4, when both groups performed under Condition 1, and in the remaining blocks, during which Group A trained under Condition 2 and Group B under Condition 1 (training protocol shown in Fig. \ref{protocol}). 
For both groups, training performance increased steadily from Day 1 to Day 3, indicating effective skill acquisition. 
On Day 4, performance levels were largely maintained, although some variability across participants was observed.

\begin{figure}[htbp]
	\centering
	\includegraphics[width=0.45\textwidth]{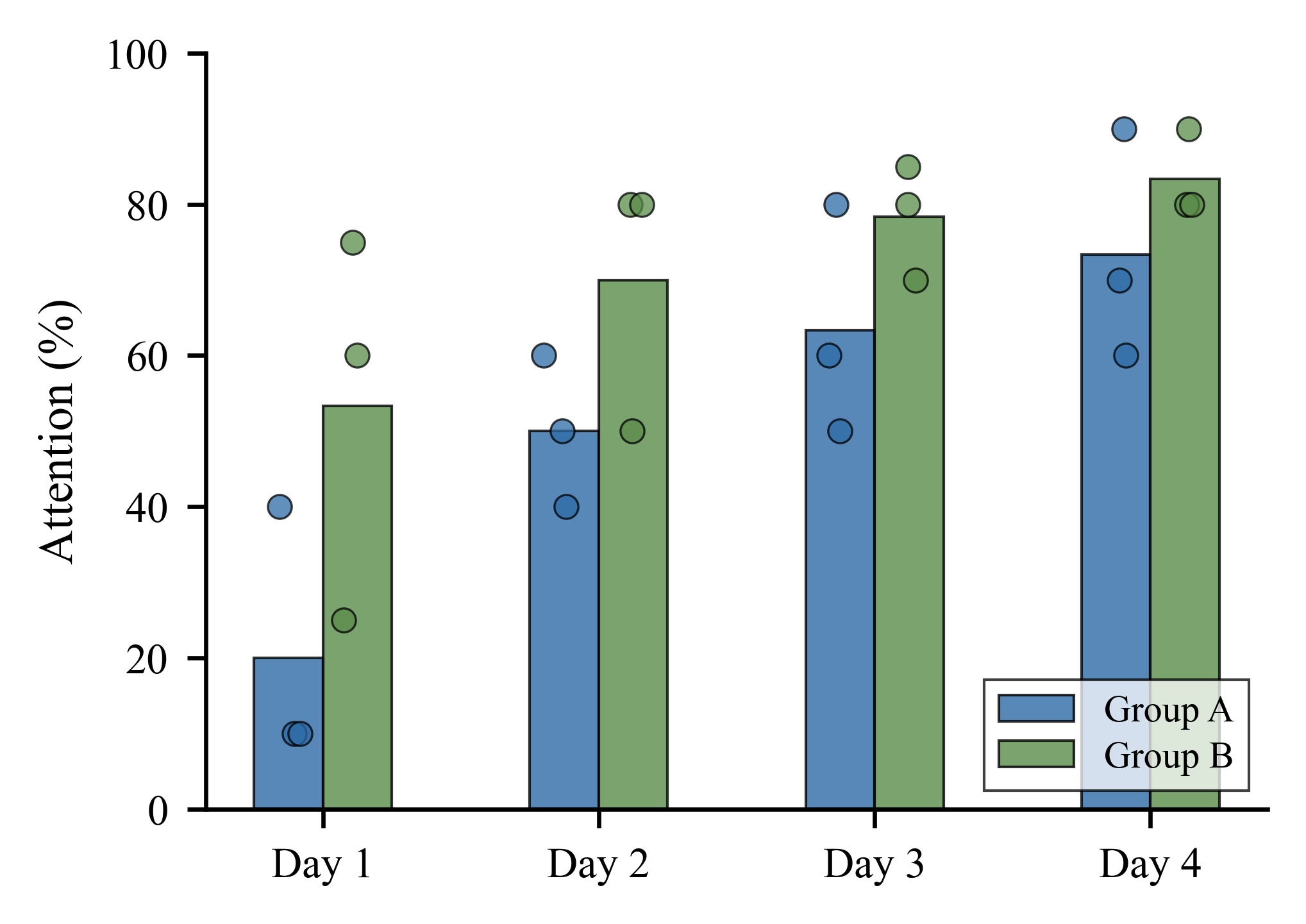}
	\caption{Attention directed toward the virtual hand across training days. 
    Bars represent the group-level mean percentages of attention to the virtual hands for Groups A and B across four training days. 
    Circles indicate individual participant data points. 
	}
	\label{attention}
\end{figure}

\begin{figure*}[htbp]
	\centering
	\includegraphics[width=0.88\textwidth]{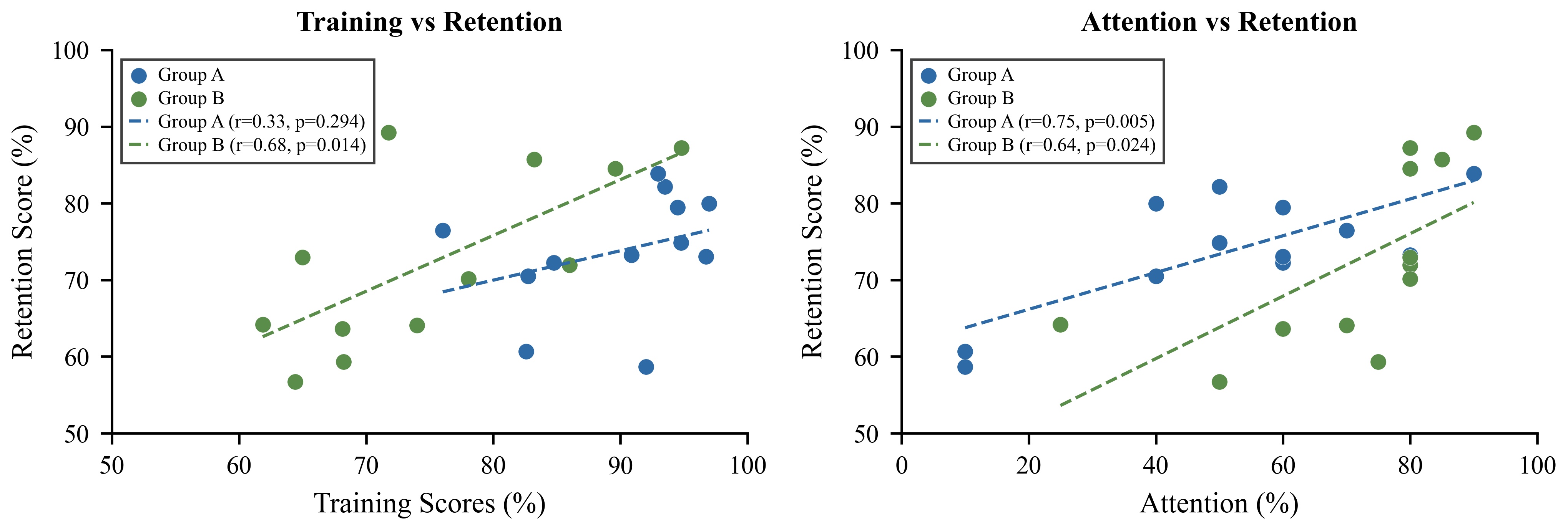}
	\caption{Correlations between retention scores and two measures: training scores (left) and attention to the virtual hand (right). Each dot represents one participant on one training day (Group A in blue; Group B in green). Dashed lines show the linear trend for each group. 
	}
	\label{correlation}
\end{figure*}

Fig. \ref{attention} illustrates the proportion of attention directed toward the virtual hand across four training days for Groups A and B. Both groups showed a progressive increase in attention from Day 1 to Day 4, reflecting a gradual shift toward an external focus with continued training. 
Group B consistently demonstrated higher attention levels than Group A across all days. The largest between-group difference occurred on Day 1 (Group A: 20\%; Group B: 53\%), with the gap narrowing over time to just 10\% by Day 4. 
Individual data points showed moderate inter-subject variability, yet both groups showed a consistent upward trend.

Figure \ref{correlation} illustrates the relationships between retention performance and training-related variables for both groups. Group B exhibited a significant positive correlation between training accuracy and retention (r = 0.684, p = 0.0142), whereas Group A showed a weaker, non-significant association (r = 0.331, p = 0.2938). In contrast, attention to the virtual hand was strongly correlated with retention in both groups: Group A (r = 0.754, p = 0.0046) and Group B (r = 0.644, p = 0.0238).

\section{Discussions}

This study introduced a virtual pre‑prosthetic training platform that integrates both muscle‑level and functional‑level feedback and evaluated its use through a proof‑of‑concept experiment that explored two FoA protocols. 
We found that (1) the platform supported the acquisition and retention of myoelectric skills across different FoA protocols; (2) the two FoA conditions showed distinct learning tendencies, suggesting complementary contributions of internal and external FoAs; and (3) external FoA appeared to play a role in how training performance related to subsequent skill acquisition. 
In short, the findings indicate that the system can be used to investigate how different FoA emphases shape myoelectric learning and provide a preliminary basis for developing training approaches that better support real‑world prosthesis use.

The design of this training platform is oriented around a common problem where training performance increases without leading to increased performance in a retention scenario \cite{stuttaford2023delaying}. 
From a motor learning perspective, an effective prosthetic skill acquisition depends not only on learning to modulate muscle contractions, an internally focused process, but also on the functional outcomes of such muscle contractions, which promotes a more external, outcome‑oriented focus \cite{parr2022scoping}.
A key innovation of the proposed platform is its ability to provide muscle- and functional-level feedback.
By integrating a V-shaped EMG cursor interface with an EMG-driven virtual hand, the system supports explicit awareness of muscle activation patterns while also presenting goal-oriented task outcomes. 
This goal‑oriented outcome is essential for myoelectric control, particularly under the abstract control scheme used in this study, where the relationship between muscle contractions and the resulting gesture is not intuitively apparent, and the sensorimotor mapping must be relearned.
Importantly, different from the existing hand model-based virtual platform \cite{boschmann2021immersive,tchimino2025effects,maas2025skill,hashim2021comparison}, the platform is able to present muscle‑centric feedback, outcome‑centric feedback, or both simultaneously, enabling systematic manipulation of internal, external, or combined FoA based on different requirements. This flexibility makes the system particularly valuable for early‑stage training, offering a basis for investigating the influence of FoA during skill acquisition.

Participants in both groups showed promise in interacting with the system and demonstrated improvements in both training and retention performance across days, suggesting the feasibility of the two FoA‑based training protocols. 
As expected, distinct patterns emerged between the two FoA conditions. Although both groups improved their training performance and gradually shifted their attention toward the virtual hand, Group A consistently showed higher training performance, whereas Group B adopted a more external FoA throughout training. 
These patterns align with the protocol design: Group A was shown the cursor in all training blocks, providing an advantage for fine‑tuning muscle modulation, while Group B trained without the cursor in most blocks, forcing participants to focus externally from the start. 

Notably, Group A initially outperformed Group B in retention scores, suggesting that Group B may not have fully consolidated its internal control representation before being directed toward an external FoA task. 
Over time, however, Group B gradually caught up and eventually showed a slight performance advantage by the final day. 
This crossover pattern indicates that both forms of FoA contribute meaningfully to learning, and that their timing may be important: 
An internal FoA, which supports the development of a stable muscle control pattern, may be more beneficial during the early stage of skill acquisition, whereas an external FoA appears to become more advantageous as training progresses.
This feature points toward the potential value of training protocols that incorporate an appropriate transition from internal to external FoA. Although only preliminarily observed in this study, such a transition may allow learners to first establish stable muscle control and then shift toward more functional performance demands, potentially benefiting users as their skills develop.

Different patterns emerged in the two groups when examining how training translated into retention performance. 
Group B showed a clearer correspondence between training accuracy and retention; however, this link was not evident in Group A.
This discrepancy shows that stable control patterns in the presence of real-time internal‑FoA‑dominant feedback do not reliably translate into retention, which is in line with previous findings \cite{stuttaford2023delaying}. 
Interestingly, external FoA was positively associated with retention scores in both groups, further indicating that external FoA cues play a critical role in consolidating skill learning \cite{wulf2013attentional} and in making training performance a predictive indicator of skill acquisition.

Several limitations should be acknowledged. As a proof-of-concept study, the sample size was limited, and the training duration was relatively short. Consequently, the results should be interpreted as preliminary and exploratory rather than definitive. 
In addition, FoA timing in the internal FoA group was not manipulated; participants could still transfer attention to the external part as training progressed. A design including separate internal-only, external-only, and mixed training conditions should be considered in future studies.
Furthermore, participants were able-bodied individuals interacting with a virtual hand model, which may not fully capture the sensory and cognitive demands experienced by individuals with limb loss. Cognitive load and attentional allocation were also not directly measured, limiting conclusions regarding the mechanisms underlying performance differences.
Future work should include larger participant cohorts, longer-term training and retention assessments, and direct measures of cognitive demand and position of FoA. Extending the platform to individuals with upper-limb amputation and integrating adaptive training protocols that dynamically adjust internal and external FoA based on user performance may further enhance clinical relevance.

\section{Conclusions}

This study introduced a virtual pre‑prosthetic training system that integrates muscle‑level and functional‑level feedback, and validated its use under two FoA protocols through a proof‑of‑concept experiment. 
The results show that early myoelectric skill learning benefits from the complementary contributions of internal and external FoA, with different FoA emphases shaping how participants adapt their control strategies over time. The integration of the external FoA may strengthen the link between training performance and subsequent skill acquisition, highlighting its relevance for developing transferable control skills.
These findings highlight the value of incorporating both forms of feedback in pre‑prosthetic training and point to the potential of this FoA‑based platform to better prepare users for real-world prosthesis use.


\bibliographystyle{IEEEtran}
\bibliography{torque}

\end{document}